\documentclass[preprintnumbers,10pt,nofootinbib]{revtex4}

\usepackage{amsmath,latexsym,amssymb,amsfonts}
\usepackage[pdftex]{color,graphicx}
\usepackage{bm}
\usepackage[normalem]{ulem}

\begin{document}


\title{Can a naked singularity be formed during the gravitational collapse of a Janis-Newman-Winicour solution? }

\author{Xiao Yan Chew$^{a}$\footnote{{\tt xiao.yan.chew@just.edu.cn}}, Il Gyeong Choi$^{b}$\footnote{{\tt eogkrrksek12@gmail.com}}, Hyuk Jung Kim\footnote{{\tt curi951007@gmail.com}} and Dong-han Yeom$^{b,c,d}$\footnote{{\tt innocent.yeom@gmail.com}}
}

\affiliation{
$^{a}$School of Science, Jiangsu University of Science and Technology, Zhenjiang 212100, China \\
$^{b}$Department of Physics Education, Pusan National University, Busan 46241, Republic of Korea\\
$^{c}$Research Center for Dielectric and Advanced Matter Physics, Pusan National University, Busan 46241, Republic of Korea\\
$^{d}$Leung Center for Cosmology and Particle Astrophysics, National Taiwan University, Taipei 10617, Taiwan
}

\begin{abstract}
The Janis-Newman-Winicour (JNW) spacetime possesses a naked singularity, although it represents an exact particle-like solution to the Einstein-Klein-Gordon theory with a massless scalar field. Here, we investigate the possible formation of a naked singularity in the JNW spacetime, using the thin-shell approximation to describe the gravitational collapse. By introducing different matter contents to construct thin-shells, we demonstrate the impossibility of naked singularity formation during the gravitational collapse unless the \textit{causality} or \textit{null energy condition} of the thin-shell is violated. Therefore, the weak cosmic censorship is satisfied even with the naked singularity of the JNW spacetime.
\end{abstract}

\maketitle

\newpage

\tableofcontents

\section{Introduction}

In general relativity (GR), a singularity is formed when an astrophysical object undergoes gravitational collapse \cite{Hawking:1973uf}. This raises the question of whether an observer that is far from the singularity can witness any event occurring in its vicinity. Assuming such observations were possible, a naked singularity would offer a fascinating opportunity to study the effects of quantum gravity. However, owing to the collapse of GR near the singularity, new physics is required for any prediction in this extreme regime.

Thus, can such solutions be obtained in GR? One simple example that can be considered is the Reissner-Nordstrom black hole, where the electric charge exceeds the mass \cite{RN}. In this case, the event horizon does not exist and the naked singularity can be observed. However, while such solutions may exist mathematically, it is still unclear if they can be generated through gravitational collapse or any other natural constructive process. Furthermore, the construction of a spacetime from gravitational collapse where the energy conditions are satisfied, while also facilitating the formation of a naked singularity, is a very challenging. Additionally, assuming a field configuration of the collapsed matter is prepared such that the asymptotic charge surpasses the asymptotic mass during gravitational collapse, the additional charges can display repulsive behaviors in complex dynamics, potentially preventing formation \cite{Hwang:2010im}.

Therefore, the emergence of a naked singularity can be prevented by two conjectures: the \textit{weak and strong cosmic censorship conjectures} \cite{Penrose:1969pc}. The weak cosmic censorship conjecture posits that a naked singularity must not be observable from gravitational collapse; thus it must be concealed by the event horizon of a black hole. Conversely, the latter posits that no observer can cross the Cauchy horizon to witness the effect of a timelike singularity. Consequently, regardless of whether the singularity is timelike or spacelike, it must be concealed by the inner apparent horizon (e.g., see \cite{Hong:2008mw}). 

However, in this study, we investigated the gravitational collapse of the Janis-Newman-Winicour (JNW) spacetime, which is an analytical solution to the Einstein-Klein-Gordon theory with a massless scalar field. Although this spacetime does not exhibit an event horizon, it displays a timelike singularity, as a consequence of bypassing the no-hair theorem\footnote{However, several classes of everywhere regularity and asymptotically flat particle-like solutions \cite{Chew:2022enh,Chew:2024bec} in the Einstein-Klein-Gordon theory can be obtained by minimally coupling the Einstein gravity and some non-positive definite scalar potentials that can violate the weak energy condition. These solutions can be smoothly connected to their corresponding hairy black holes in the small horizon limit \cite{Chew:2022enh,Chew:2023olq}.}\cite{Israel:1967wq}. Historically, Fisher \cite{Fisher:1948yn}, Buchdahl, Bergman, and Leipnik \cite{Bergmann:1957zza} obtained this solution, which was subsequently rediscovered by Janis, Newman, and Winicour \cite{Janis:1968zz}. Thereafter, several extended solutions for the JNW spacetime have been investigated. For instance, the spacetime can take the form of a Vaidya-like metric which is adopted for the constructions of naked singularity solutions from dynamical gravitational collapses \cite{Roberts:1989sk}. The author reported that regular solutions for gravitational collapses were allowed, although these solutions required an additional \textit{ad hoc} matter field or a matter shell at the Minkowski--naked singularity solution junction. Moreover, such an additional matter field must be justified to guarantee the consistency of the dynamics. Furthermore, to connect from Minkowski to the naked singularity solution, the dynamics of a \textit{thin-shell} at the junction surface, must be carefully resolved. However, the literature does not contain detailed investigations of thin-shell dynamics.

Furthermore, as the aim of our study was to investigate if the JNW spacetime can be a counterexample of the weak cosmic censorship conjecture, we adopted the \textit{thin-shell approximation} \cite{Israel:1966rt}. As this approximation can integrate the dynamical back-reactions of the metric, and treat a realistic energy condition in the shell, it was considered a more realistic and controllable method for investigating gravitational collapses. In our attempt to violate the weak cosmic censorship \cite{Roberts:1989sk}, we need to introduce a delta-function shell anyway. Here, we consider that the thin-shell has to satisfy the energy conditions. In principle, it can be constructed using various matters satisfying energy conditions \cite{Chen:2015lbp}. Suppose we cannot find a successful gravitational collapse solution from the thin-shell approximation by varying the matter configuration of the shell; in that case, this will be strong evidence that weak cosmic censorship is physically sustainable even with the JNW solution possessing a naked singularity.

The rest of this paper is organized as follows. Sec.~\ref{sec:JNW} discusses the JNW analytic solution and the junction equation between the Minkowski or pure anti-de Sitter(AdS) solution and the JNW solution. Sec.~\ref{sec:int}, presents the numerical and analytical ways. Finally, Sec.~\ref{sec:dis} summarizes the study and recommends possible future outlooks.

\section{\label{sec:JNW}Model}

\subsection{Janis-Newman-Winicour spacetime}

The JNW spacetime describes a static, stationary, asymptotically flat, and spherically symmetric spacetime exhibiting a naked singularity \cite{Janis:1968zz}. It can be obtained analytically by solving the equations of motion:
\begin{equation}
    R_{\mu \nu} = 2 \nabla_\mu \varphi \nabla_\nu \varphi \,, \quad  \nabla_\mu \nabla^\mu \varphi = 0 \,,
\end{equation}
which can be derived from the following theory
\begin{equation}
    S = \int d^4 x \left( \frac{R}{16\pi G} - \frac{1}{2} \nabla^\mu \varphi \nabla_\mu \varphi   \right) \,.
\end{equation}
The explicit form of the JNW spacetime is given by
\begin{eqnarray} \label{JNW1}
ds^{2} = - g(R) dt^{2} + \frac{1}{g(R)} dR^{2} + r^{2} d\Omega^{2},
\end{eqnarray}
where
\begin{eqnarray}
g(R) &=& \left( \frac{2R - r_{0} (\mu-1)}{2R + r_{0} (\mu+1)} \right)^{1/\mu}, \\
r^{2} &=& \frac{1}{4} \left( 2R + r_{0} (\mu+1) \right)^{1+1/\mu} \left( 2R - r_{0} (\mu-1) \right)^{1-1/\mu}, \\
\varphi(R) &=& \frac{A}{\mu} \ln \left| \frac{2R - r_{0} (\mu-1)}{2R + r_{0} (\mu+1)} \right|,\\
\mu &\equiv& \left( 1 + \frac{4\kappa A^{2}}{r_{0}^{2}} \right) \geq 1,\\
\kappa &\equiv& 8\pi,\\
r_{0} &\equiv& 2m,
\end{eqnarray}
where $m$ is the Arnowitt-Deser-Misner mass, $A$ is the scalar charge of the scalar field, and $d\Omega^{2}$ is a two-dimensional sphere $S^2$. The location of the naked singularity is determined by condition $g(R_{\text{sing}}) = 0$, which is equivalent to
\begin{equation}
R_{\text{sing}} = \frac{r_0}{2} (\mu-1) = \frac{\kappa A^2}{m} \,.
\end{equation}
Note that the scalar field $\varphi$ also diverges exactly at $R_{\text{sing}}$.

\subsection{Thin-shell formalism}

A thin-shell spacetime with manifold $\mathcal{M}$ is formed by joining two distinct spacetimes $(\mathcal{M}=\mathcal{M}_+ \cup \mathcal{M}_-)$ at the boundary $\Sigma$ where $\Sigma =\Sigma_+ + \Sigma_-$ comprises manifolds $\mathcal{M}_+$ and $\mathcal{M}_-$ with boundaries $\Sigma_+$ and $\Sigma_-$, respectively; the metrics are given by $g^+_{\mu \nu}(x^\mu_+)$ and $g^-_{\mu \nu}(x^\mu_-)$ regarding independent coordinate systems $x^\mu_+$ and $x^\mu_-$, respectively \cite{Garcia:2011aa}. We employ the following metric to describe $\mathcal{M}_\pm$ of the thin-shell spacetime: 
\begin{eqnarray} \label{ts}
ds_{\pm}^{2} = - f_{\pm}(r) e^{2\Phi_{\pm}(r)} dt^{2} + \frac{1}{f_{\pm}(r)} dr^{2} + r^{2} d\Omega^{2},
\end{eqnarray}
where subscripts $+$ and $-$ represent the outer and inner parts of the thin-shell, respectively. Here, we assume the JNW and Minkowski or Ads spacetime to be $\mathcal{M}_+$ and $\mathcal{M}_-$ respectively, as follows:
\begin{eqnarray}
\Phi_{-} &=& 0,\\
f_{-} &=& 1 + \frac{r^{2}}{\ell^{2}}.
\end{eqnarray}
As the metric of the JNW spacetime (Eq.~\eqref{JNW1}) does not match exactly with our metric (Eq.~\eqref{ts}), we rewrite Eq.~\eqref{JNW1} by expressing function $R$ as the function of the physical areal radius $r$, as follows:
\begin{eqnarray}
ds^{2} &=& - g(R) dt^{2} + \frac{1}{g(R)} \left( \frac{dR}{dr} \right)^{2} dr^{2} + r^{2} d\Omega^{2} \,, \\
&=& - f_+(r) e^{2\Phi_+(r)} dt^{2} + \frac{1}{f_+(r)} dr^{2} + r^{2} d\Omega^{2} \,. \label{JNW3}
\end{eqnarray}
Therefore, the comparison of both metrics above yields the following relations,
\begin{eqnarray}
\Phi_+(r) &\equiv& \ln \left( \frac{dR}{dr} \right) \,, \\
f_+(r) &\equiv& g(R) \left( \frac{dR}{dr} \right)^{-2} \,.
\end{eqnarray}
Note that we omit the notation $r_\pm$, using only $r$ to represent them.

Next, we examine the junction condition at $\Sigma$ to study the collapse of a timelike thin-shell. As the thin-shell spacetime is static and spherically symmetric, the hypersurface at $\Sigma$ can be described by the following induced metric with coordinate $x^\mu(\tau,\theta,\phi)=(t(\tau),r(\tau),\theta,\phi)$, as follows:
\begin{eqnarray}
ds^{2}_{\mathrm{shell}} = - d\tau^{2} + r^{2}(\tau) d\Omega^{2}\,,
\end{eqnarray}
where $\tau$ is the proper time of an observer comoving with the junction surface. Moreover, the Israel junction condition yields the following relations to describe $\Sigma$ \cite{Garcia:2011aa,Israel:1966rt}:
\begin{eqnarray}
\epsilon_{-} \sqrt{f_{-}(r) + \dot{r}^{2}} - \epsilon_{+} \sqrt{f_{+}(r) + \dot{r}^{2}} &=& 4\pi \sigma(r) r \,, \label{jc1} \\
\lambda(r) &=& - \frac{r}{2} \left(\sigma'(r) - \Xi(r) \right) - \sigma(r) \,, \label{jc2}
\end{eqnarray}
where $\sigma(r)$, and $\lambda(r)$ are the tension and pressure at $\Sigma$, respectively. Additionally, $\epsilon_\pm$ is $+1$ ($-1$) if $r$ increases (decreases) along the outward normal direction, and the superscript prime denotes the derivative of a function relating to $r$. We define
\begin{eqnarray}
\Xi(r) \equiv \frac{1}{4\pi r} \left( \epsilon_{+} \Phi_{+}'(r) \sqrt{f_{+}(r) + \dot{r}^{2}} - \epsilon_{-} \Phi_{-}'(r) \sqrt{f_{-}(r) + \dot{r}^{2}} \right)\,,
\end{eqnarray}
where $\dot{r}$ is the derivative of a function relating to $\tau$. Eq.~\eqref{jc1} could be recast into a simpler form:
\begin{eqnarray}
\dot{r}^{2} + V_{\mathrm{eff}}(r) = 0,\, \label{jc1a}
\end{eqnarray}
where
\begin{eqnarray}
V_{\mathrm{eff}}(r) \equiv f_{+}(r) - \frac{\left(f_{-}(r) - f_{+}(r) - 16\pi^{2}\sigma^{2}(r)r^{2}\right)^{2}}{64\pi^{2} \sigma^{2}(r) r^{2}}\,.
\end{eqnarray}
Note that the solution to $\dot{r}$ can only be obtained when $V_{\mathrm{eff}}(r) < 0$. However, analyzing Eqs.~\eqref{jc1a} in the radial coordinate $R$ rather than $r$ is more convenient. Therefore, we rewrite the above equation, as follows:
\begin{eqnarray}
\dot{R}^{2} \left(\frac{dr}{dR}\right)^{2} + V_{\mathrm{eff}}(R) = 0\,, \label{jc1c}
\end{eqnarray}
where 
\begin{eqnarray}\label{eq:effpot}
V_{\mathrm{eff}}(R) \equiv f_{+}(R) - \frac{\left(f_{-}(R) - f_{+}(R) - 16\pi^{2}\sigma^{2}(R)r^{2}(R)\right)^{2}}{64\pi^{2} \sigma^{2}(R) r^{2}(R)}.
\end{eqnarray}
Similarly, the form of Eq.~\eqref{jc2} in coordinate $R$ is given by 
\begin{equation}
\lambda(R) = - \frac{r(R)}{2} \left( \frac{d\sigma}{dR} \frac{dR}{dr} - \frac{1}{4\pi r(R)} \left( \epsilon_{+} \frac{d\Phi_{+}}{dR} \frac{dR}{dr} \sqrt{f_{+}(R) - V_{\mathrm{eff}}(R)} - \epsilon_{-} \frac{d\Phi_{-}}{dR} \frac{dR}{dr} \sqrt{f_{-}(R) - V_{\mathrm{eff}}(R)} \right) \right) - \sigma(R) \,. \label{jc2a}
\end{equation}

To support the junction $\Sigma$, it is necessary to incorporate physical matter, represented by the equation of state $w(R) \equiv \lambda/\sigma$, which must satisfy the null energy condition in order to guarantee that $w \geq -1$ \cite{Chen:2015lbp}\footnote{Generally, the tension $\sigma$ must explicitly include the contribution from $\varphi$ as $w(R) \equiv \lambda/\sigma + P_{\text{scalar}}$, where $P_{\text{scalar}}$ is the pressure of $\varphi$. However, we assume that $\lambda/\sigma$ has already included $P_\text{scalar}$ for simplicity.}. Recall that the inner shell is assumed to be a Minkowski or AdS spacetime; therefore, we set $\Phi_{-}' = 0$ in Eq.~\eqref{jc2a} since $\Phi_-=0$. Thereafter, we introduced $X(R) \equiv \sigma^{2}(R)$ to further rewrite Eq.~\eqref{jc2a} as 
\begin{eqnarray}\label{eq:X}
X'(R) + \left( \frac{4(w+1)}{r(R)} \frac{dr}{dR} + \epsilon_{+} \frac{d\Phi_{+}(R)}{dR} \right) X(R) = \epsilon_{+} \frac{d\Phi_{+}(R)}{dR} \frac{(f_{-}(R) - f_{+}(R))}{16 \pi^{2} r^{2}(R)} \,.
\end{eqnarray}
The condition $X > 0$ must be satisfied since $\sigma=\sqrt{X}$. Thus, we need to consider this condition and simultaneously solve Eqs.~\eqref{eq:effpot} and \eqref{eq:X} by integrating them from large $R$ $(R \rightarrow \infty)$ backward to a finite value of $R$. Hence, the initial condition of $X$ at a large $R$ was $X(\infty)=X_0$, which is the asymptotic value of the tension, i.e., $\sigma^2_0=X_0$.

\section{\label{sec:int}Analyses of the solutions}

\subsection{General analyses using the effective potential}

\begin{figure}
\centering
\includegraphics[scale=0.7]{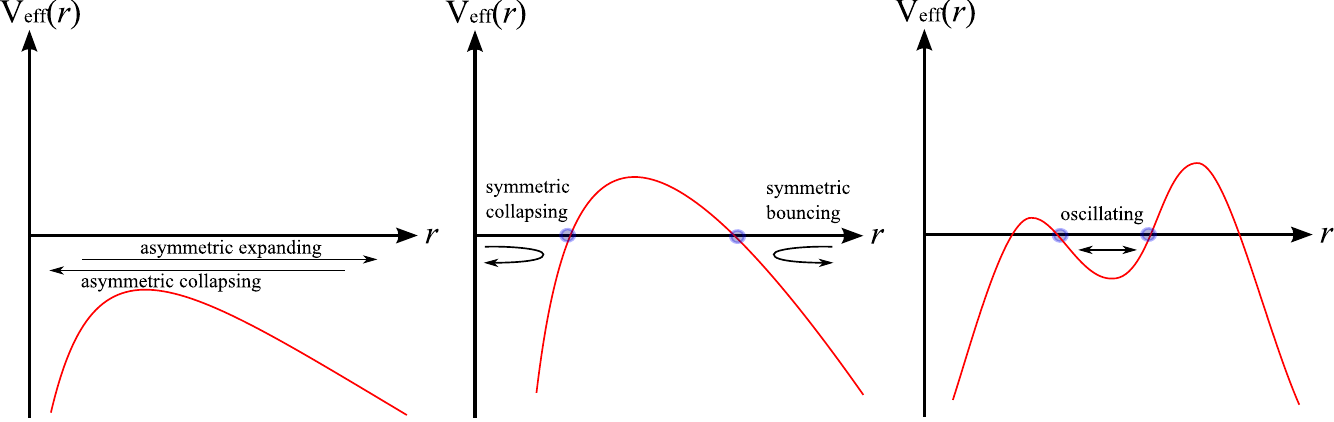}
\caption{\label{fig:pots}Classification of the solutions for Eq.~\eqref{jc1c}: asymmetric expansion or collapse (left), symmetric expansion or bouncing (middle), or oscillating (right)solutions are allowed.}
\end{figure}

The profile of a solution for Eq.~\eqref{jc1c} highly depends on that of $V_{\mathrm{eff}}(r)$. Generally, three distinct types of solutions may arise for Eq.~\eqref{jc1c} when $V_{\mathrm{eff}}  < 0$. As illustrated in Fig.~\ref{fig:pots} (left), the first type of solution may exhibit asymmetric collapse or expansion when $V_{\mathrm{eff}}(r) < 0$, without containing any zeros for $r>0$ or $R>0$. In this case, a shell just moves directly from spatial infinity to a naked singularity or vice versa \cite{Freivogel:2005qh}. Conversely, Fig.~\ref{fig:pots} (middle) illustrates the second type of solutions, which can manifest as a symmetric collapse when $V_{\mathrm{eff}}(r)  < 0$ with $r$ less than the smallest root or symmetric bouncing when $V_{\mathrm{eff}}(r)  < 0$ with $r$ larger than the largest root. This indicates that a shell can collapse (expand) from spatial infinity (a naked singularity) to a minimum radius (the spatial infinity), where both processes terminate at a certain radius, exhibiting an analogous bouncing effect \cite{Freivogel:2005qh}. Finally, Fig.~\ref{fig:pots} (right) illustrates the third solution type, which can oscillate between the two roots of $V_{\mathrm{eff}}(r)  < 0$, is enclosed by them \cite{Chen:2015lbp}.

\begin{figure}
\centering
\includegraphics[scale=0.7]{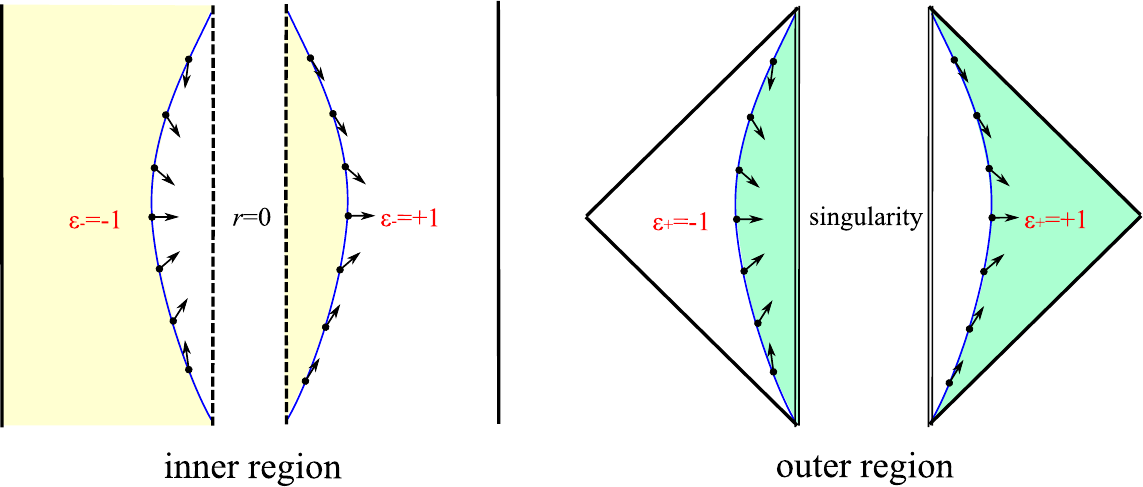}
\caption{\label{fig:signs}Interpretation of the Penrose diagram for the inside (left) and outside (right) symmetric collapsing shells; the inner and outer geometries were assumed to be the Ads and JNW solutions, respectively.}
\end{figure}

Among these three scenarios, only the second solution can behave analogously with the collapse of an astrophysical object because the matter inside a star typically exists within a finite $R$. Fig.~\ref{fig:signs} describes a typical timelike symmetric collapsing shell; the left and right figures correspond to the inner (Minkowski or Ads solution) and outer (JNW solution) regions, respectively. Furthermore, the left part of the inner region (yellow-colored region) and the right part of the outer region (green-colored region) are cut and pasted. From the inner to outer regions, we determine the outward normal direction (arrows in Fig.~\ref{fig:signs}), if $r$ increases along the outward normal direction respectively. We choose $\epsilon_{\pm} = + 1$, as it represents a physically sensible gravitational collapsing process. This condition can also confirmed by the following extrinsic curvatures:
\begin{eqnarray}
\beta_{\pm}(R) = \frac{f_{-}(R) - f_{+}(R) \mp 16 \pi^{2} \sigma^{2}(R) r^{2}(R)}{8\pi \sigma(R) r(R)} = \epsilon_{\pm} \sqrt{\dot{r}^{2} + f_{\pm}(R)} \,,
\end{eqnarray}
where $\epsilon_{+} = \textrm{sign}~\,\beta_{+}$ that must be substituted into Eq.~(\ref{eq:X}).

\subsection{Classification and interpretation of the solutions}

By numerically solving Eq.~\eqref{eq:X}, we only manage to obtain two different types of $X$ by tuning some parameters such as $A$, $m$, $l$, $\sigma_0$, and $w$. The first type solution can be obtained when the tension of the shell vanishes at a finite radius, i.e., $X(R)=0$. The second type solution can be obtained when the tension of the shell diverges at the singularity. We demonstrate both cases in the following subsections and discusse whether or not they could violate the cosmic censorship conjecture.

\begin{figure}
\centering
\mbox{
(a)
\includegraphics[scale=0.6]{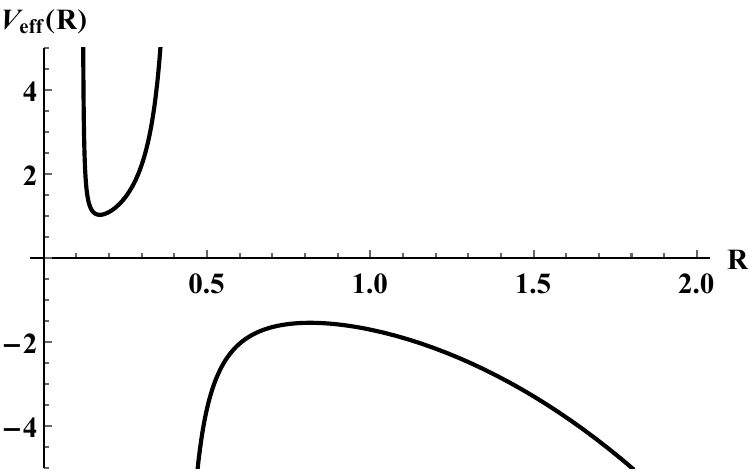}
(b)
\includegraphics[scale=0.6]{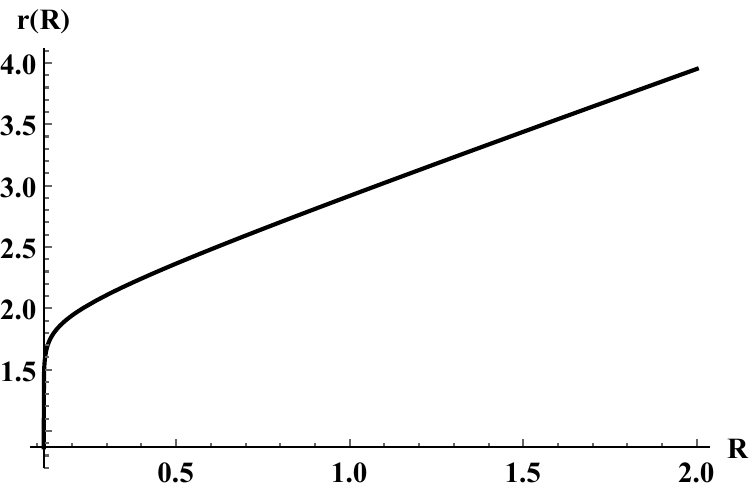}
}
\mbox{
(c)
\includegraphics[scale=0.6]{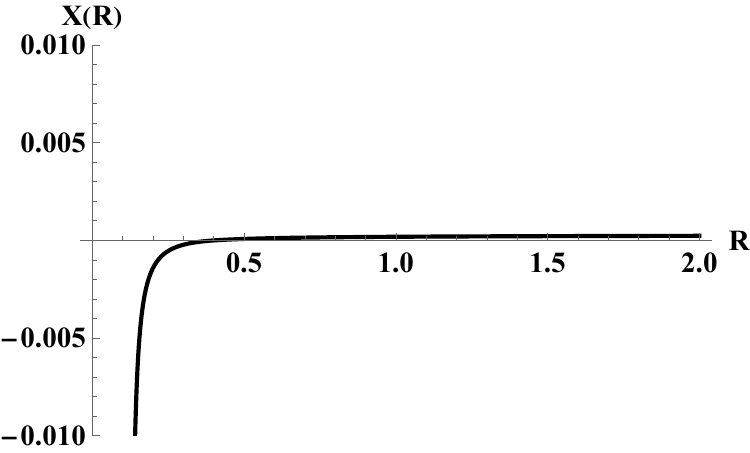}
}
\caption{\label{fig:type1} Solutions of the thin shell approximation with vanishing tension at a finite radius with $A = 0.1$, $m = 1$, $\ell = 1$, $\sigma_{0} = 0.015$, and $w = -1$: (a) $V_{\mathrm{eff}}(R)$, (b) $r(R)$, and (c) $X(R)$.}
\end{figure}

\begin{figure}
\centering
\mbox{
(a)
\includegraphics[scale=0.6]{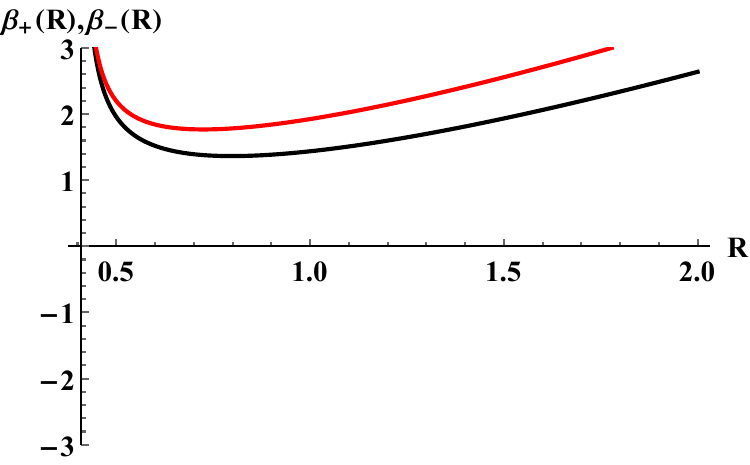}
(b)
\includegraphics[scale=0.6]{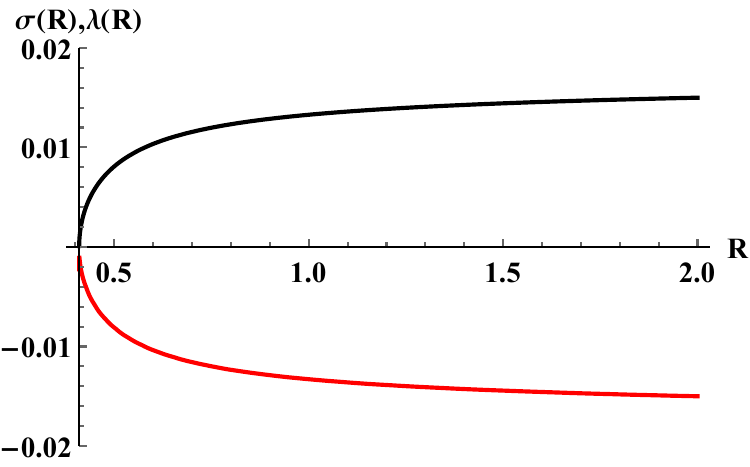}
}
\caption{\label{fig:type1_2} Other properties of the thin shell with $A = 0.1$, $m = 1$, $\ell = 1$, $\sigma_{0} = 0.015$, and $w = -1$: (a) $\beta_{+}(R)$ (black), $\beta_{-}(R)$ (red) and (b) $\sigma(R)$ (black), $\lambda(R)$ (red).}
\end{figure}

\subsubsection{Type 1: Timelike to spacelike bending}

Fig.~\ref{fig:type1} (a) shows the $V_{\mathrm{eff}}(R)$ profile of the shell with $A = 0.1$, $m = 1$, $\ell = 1$, $\sigma_{0} = 0.015$, and $w = -1$, corresponding to the asymmetric collapse without any zeros (Fig.~\ref{fig:pots} (left)). Hence, all the solutions are only valid when $V_{\mathrm{eff}}(R)<0$ for the $R$ in $(0.4096,\infty)$ range. Figs.~\ref{fig:type1} (b) and (c) show the $r(R)$ and $X(R)$ profiles, respectively; they decrease monotonically from a large $R$ to a finite radius ($R \sim 0.4096$) when $X(R)=0$, resulting in the divergence of $r$ at $R \sim 0.4096$. Notably, that the naked singularity of the JNW spacetime is less than $R \sim 0.4096$. 

Furthermore, Fig.~\ref{fig:type1_2} (a) reveals that the extrinsic curvatures ($\beta_{\pm}$) are positive definite for the range of $R$ in $(0.4096,\infty)$. This should be expected, as it describes a consistent gravitational collapse process. Fig.~\ref{fig:type1_2} (b) shows that the tension $\sigma(R)$ (black) decreases monotonically to zero at the critical radius ($R \sim 0.4096$), whereas the pressure $\lambda(R)$ (red) increases monotonically to zero at $R \sim 0.4096$. We confirm that the ratio ($\lambda(R)/\sigma(R)$) can satisfy the relation $w=\lambda(R)/\sigma(R) = -1$, demonstrating that the null energy condition is always satisfied. 

Therefore, what could be the physical interpretation of the solution when $X<0$. In this scenario, the shell has reached a critical radius and can be transformed into a \textit{null} shell within a finite proper time. Thus, we can directly redefine the function $\sigma$ by the following expression:
\begin{equation}
    \sigma \rightarrow - i \sigma.
\end{equation}
Based on this, we can also redefine some other functions, as follows:
\begin{eqnarray}
\lambda &\rightarrow& + i \lambda, \\
d\tau &\rightarrow& -ids, \\
\sqrt{ W } &\rightarrow& - i \sqrt{ W },
\end{eqnarray}
where $W$ is an arbitrary function and the last equation is necessary to ensure the consistency of Eq.~\eqref{eq:X}. Although the quantity ($ds$) has become spacelike, it could still effectively describe the dynamics of the spacelike shell for $R \lesssim 0.4096$ (Fig.~\ref{fig:type1} (a)) using the following equation \cite{Brahma:2018cgr}:
\begin{eqnarray}
\left(\frac{dr}{ds} \right)^{2} - V_{\mathrm{eff}}(R) = 0.
\end{eqnarray}

\begin{figure}
\centering
\includegraphics[scale=0.7]{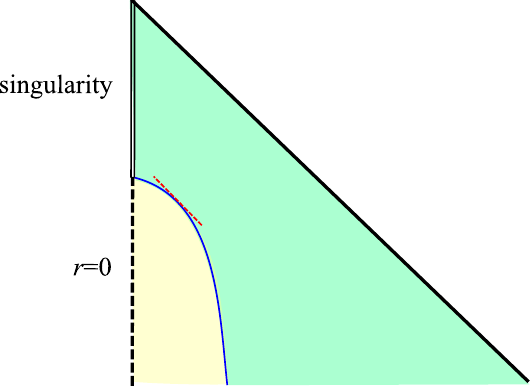}
\caption{\label{fig:type1_Pen}Interpretation of the type of vanishing tension. The shell approaches a null direction (red dashed line) within a finite time, eventually becoming a spacelike shell, though this violates the causality principle.}
\end{figure}

Therefore, our solution can describe the collapse of the shell which could eventually touch the singularity of the JNW solution (Fig.~\ref{fig:type1_Pen}) in response to the timelike-to-spacelike transition. In principle, this transition is mathematically possible, as demonstrated by \cite{Chen:2017pkl}. However, one might question can we physically accept the violation of causality as the consequence of the violation of weak cosmic censorship.

\subsubsection{Type 2: Diverging tensions}

\begin{figure}
\centering
\mbox{
(a)
\includegraphics[scale=0.6]{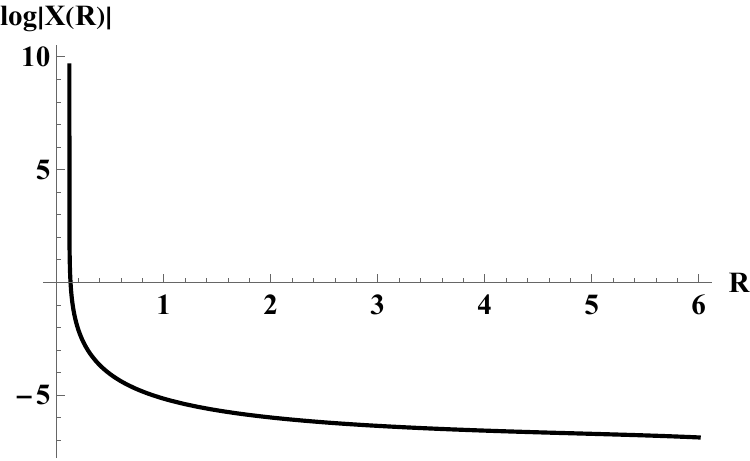}
(b)
\includegraphics[scale=0.6]{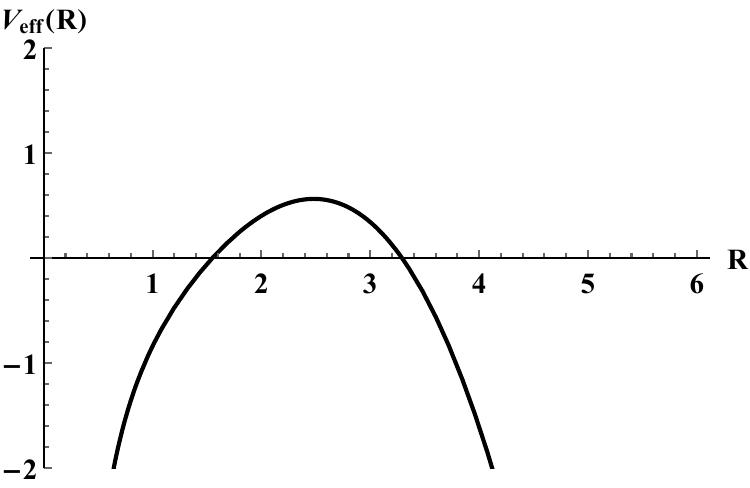}
}
\mbox{
(c)
\includegraphics[scale=0.6]{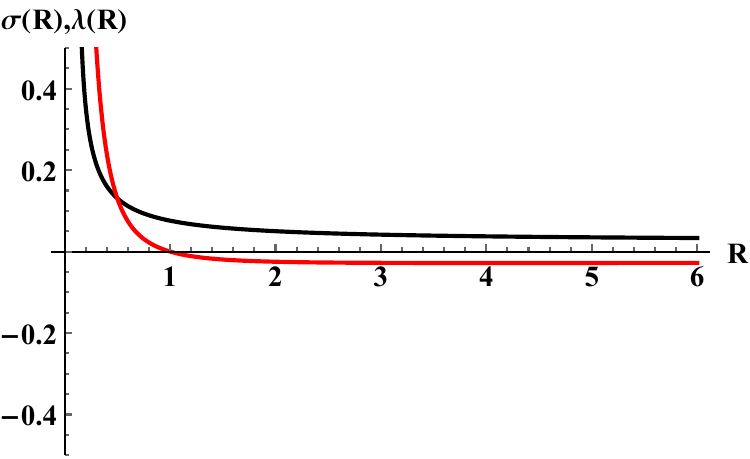}
(d)
\includegraphics[scale=0.6]{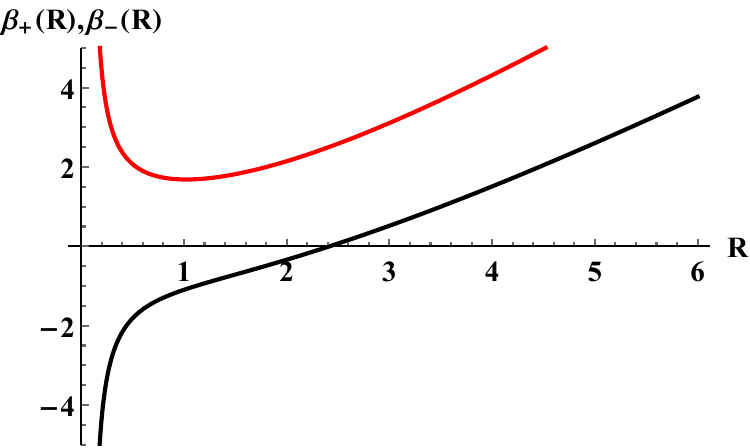}
}
\caption{\label{fig:type2}Example of diverging tension at the singularity: (a) $\log |X|$, (b) $V_{\mathrm{eff}}(R)$, (c) $\sigma(R)$ (black), $\lambda(R)$ (red), and (d) $\beta_{+}$ (black), $\beta_{-}$ (red). Here, $A = 0.1$, $m = 1$, $\ell = 1$, $\sigma_{0} = 0.05$, and $w = -1 + 1/R$.} 
\end{figure}

By tuning some parameters, we obtain another solution for $X$, which does not vanish at a finite radius but diverges near the singularity of the JNW solution (Fig.~\ref{fig:type2} (a)). This solution is obtained when the $V_{\mathrm{eff}}(R)$ profile could describe the symmetric collapse (Fig.~\ref{fig:type2} (b)).

Moreover, Fig.~\ref{fig:type2} (c) shows that $\sigma(R)$ (black) decreases monotonically as $R$ increases, approaching a constant. However, Fig.~\ref{fig:type2} (d) shows that $\beta_{+}$ (black) could become negative for the small $R$ regime. Therefore, the symmetric collapsing solution corresponds to $\epsilon_{+} = - 1$, which is not physically sensible for describing gravitational collapses.

Is this problem unavoidable or not.? If the tension diverges near the singularity, $f_{-} - f_{+} - 16\pi^{2} \sigma^{2} r^{2} \simeq - 16\pi^{2} \sigma^{2} r^{2} < 0$. Therefore, a point exists such that
\begin{eqnarray}
f_{-} - f_{+} - 16\pi^{2} \sigma^{2} r^{2} = 0
\end{eqnarray}
is satisfied. At this point, $\beta_{+}$ becomes zero, and $\beta_{+}$ changes from positive to negative as $R$ decreases (e.g., $R \sim 2.4172$; right side of Fig.~\ref{fig:type2}). At this moment, $V_{\mathrm{eff}} = f_{+}$ is satisfied, whereas $f_{+}$ is a positive definite in all parameter regimes. Therefore, we conclude that a point where $\beta_{+} = 0$ is satisfied exists as $\sigma$ diverges. At this point, $V_{\mathrm{eff}} > 0$ (e.g., $R \sim 2.4172$; left side of Fig.~\ref{fig:type2}); this corresponds to the effective potential barrier. This ensures that the corresponding $\beta_{+}$ is negative in the small $R$ regime even though the effective potential can be negative, explaining why we cannot obtain a consistent collapsing solution with diverging tension.

\subsection{Is it possible to obtain another solution for $X$?}

Thus far, we have only obtained two shell types for $X$, where the first type exhibits a vanishing tension at a finite radius which could violate the causality principle; the other shell does not describe a physical collapse, as it exhibits a diverging tension at the singularity with the negative extrinsic curvature.  

Nevertheless, we might wonder if there is any possibility of obtaining other solutions. Although we do not offer a mathematical proof, we provide some reasonable arguments to support our findings by performing some simple analysis on Eq.~\eqref{eq:X} for the two scenarios, as follows: 
\begin{itemize}
\item[--] If $X$ is sufficiently small and $\epsilon_{+} = + 1$, we can approximate Eq.~\eqref{eq:X} as
\begin{eqnarray}
X'(R) \simeq \epsilon_{+} \frac{d\Phi_{+}(R)}{dR} \frac{(f_{-}(R) - f_{+}(R))}{16 \pi^{2} r^{2}(R)},
\end{eqnarray}
where $f_{-} - f_{+} > 16\pi^{2} \sigma^{2} r^{2} > 0$; hence, the right-hand side is positive definite. Within this limit, $X \simeq \alpha R + \beta$ may be approximately written with constants $\alpha > 0$ and $\beta$. Therefore, $X$ must decrease with $R$. This indicates that $X$ becomes zero at a non-vanishing $R$.
\item[--] However, if $X$ is sufficiently large, we can approximate that
\begin{eqnarray}
\frac{X'(R)}{X(R)} \simeq - \left( \frac{4(w+1)}{r(R)} \frac{dr}{dR} + \epsilon_{+} \frac{d\Phi_{+}(R)}{dR} \right).
\end{eqnarray}
In the small $R$ regime, $dr/dR \gg 1$ (middle of Fig.~\ref{fig:type1}); thus, the right-hand side is negative definite, as long as we assume $w + 1 > 0$. From this, $X$ represents an exponentially decreasing function as $R$ increases; put differently, $X$ increases exponentially as $R$ decreases. Hence, we cannot avoid the tension divergence as $R$ approaches the singularity.
\end{itemize}
Notably, a clear loophole exists in the arguments. If $w + 1 < 0$ or one \textit{violates the null energy condition}, we expect that $X$ will never diverge for a small $R$ limit, and allow for the violation of cosmic censorship.

Therefore, in conclusion, to violate the cosmic censorship conjecture based on the JNW solution, we must violate the \textit{causality} or \textit{null energy condition}. This conclusion appears to be consistent with the other cosmic censorship conjecture.

\section{\label{sec:dis}Discussion}

In this study, we investigate whether the JNW naked singularity solution could be obtained from a reasonable model of gravitational collapse. To induce gravitational collapse, we introduce a thin-shell junction that connects the inner Minkowski or AdS and outer JNW solutions. Assuming the shell can touch the singularity with a reasonable shell-energy condition, we can confirm the possibility of violating weak cosmic censorship.

However, we obtain a negative conclusion regarding the possibility of violating the cosmic censorship conjecture. As the tension tends to diverge when the shell dynamics is well-defined for all domains, we cannot obtain a reasonable solution because the extrinsic curvature must be negative near the singularity. If we consider the initial condition with the positive extrinsic curvature that induces gravitational collapse (i.e., if we force the collapse), the shell would approach the null direction in a finite time with a non-zero radius. Hence, the shell might exhibit acausal dynamics, which is physically inconsistent.

Therefore, we conclude that the violation of the cosmic censorship conjecture for the JNW solution is impossible unless the null energy condition or causality principle is violated. However, we further inquire about the meaning of the causality violation. If the shell is \textit{forced} to collapse, it would approach the null direction within a finite proper time, although this would not imply that the shell would physically evolve to the null or spacelike direction. Such dynamics appear because we assume the thin-shell approximation; hence, a rather physically reasonable interpretation is that \textit{the consistency of the thin-shell approximation would disintegrate within a finite time}. For a charged black hole case, the formation of the naked singularity is not allowed because of the electric repulsion; similarly, the JNW naked singularity cannot be formed from the gravitational collapses, as the shell might exert repulsive effects that violate the thin-shell approximation.

So far, our computations concentrate on the JNW solution. However, this analysis could be generalized to other classes of solutions with naked singularity. The trade-off between the gravitational collapses and causality violation might represent a universal behavior of gravitational collapses with naked singularity models. We would consider this interesting topic as a future research topic.

\newpage

\section*{Acknowledgment}

DY was supported by the National Research Foundation of Korea (Grant No. : 2021R1C1C1008622, 2021R1A4A5031460). XYC is supported by the starting grant of Jiangsu University of Science and Technology (JUST).

\end{document}